# An Application of the Mobile Transient Internet Architecture to IP Mobility and Inter-Operability

Joud Khoury and Nicolas Nehme-Antoun
and Chaouki Abdallah
Electrical and Computer Engineering Department
University of New Mexico
Albuquerque, NM 87131
{jkhoury, nicolas, chaouki}@ece.unm.edu

Henry N Jerez
Senior Research Scientist
Corporation for National Research Initiatives
Reston, VA 20191
hjerez@cnri.reston.va.us

***Abstract*— We introduce an application of a mobile transient network architecture on top of the current Internet. This paper is an application extension to a conceptual mobile network architecture. It attempts to specifically reinforce some of the powerful notions exposed by the architecture from an application perspective. Of these notions, we explore the network expansion layer, an overlay of components and services, that enables a persistent identification network and other required services. The overlay abstraction introduces several benefits of which mobility and communication across heterogenous network structures are of interest to this paper. We present implementations of several components and protocols including gateways, Agents and the Open Device Access Protocol. Our present identification network implementation exploits the current implementation of the Handle System through the use of distributed, global and persistent identifiers called *handles*. Handles are used to identify and locate devices and services abstracting any physical location or network association from the communicating ends. A communication framework is finally demonstrated that would allow for mobile devices on the public Internet to have persistent identifiers and thus be persistently accessible either directly or indirectly. This application expands IP inter-operability beyond its current boundaries.**

## I. Introduction

The current Internet implementation is inherently based on a location and association aware communication substrate. Virtual circuits must be established between communicating endpoints in the form of connections that are bound to physical attachment points (IP addresses), thus hindering the evolution of the network. Mobility, transient communication and persistent identification are some basic examples of essential needs of current networks that are rendered impossible with the present Internet implementation. These features should be inherent characteristics of any future proposed network architecture that wishes to set the ground for future research. To overcome the current limitations, we are working on a new transient architecture destined to merge existing and future networks by introducing a new approach to identification and data delivery. The architecture aims at enabling seamless end-to-end communication between mobile devices over a transient substrate and across heterogeneous network infrastructures. Persistent identification is a major notion that the architecture promotes. It suggests implementing a persistent identification network on top of a distributed overlay that is an aggregation of a globally coordinating set of gateways and agents. The architecture identifies all network entities with global and persistent identifiers that are location and association independent. Mobile Entities (MEs), which are abstractions of actual network components, could be explicit entities of which end hosts, services, processes and users are the most common. Research also refers to these elements as network endpoints. These entities could also be Implicit MEs, which include objects that are mobile in nature, for example, traffic, content, digital objects and agents. The architecture abstracts both types of MEs into Digital Entities [1] that are uniquely and persistently identified. As far as this paper is concerned, persistent identification is offered by the Handle System [1], [2], [3], [4].

The Handle System is a distributed system extensively used as an indirection layer for the management of persistent identifiers. It has most of the required characteristics in terms of security, scalability and reliability to identify digital entities fostering inter-operability among heterogeneous networks.

This paper explores the network IP layer expansion, that enables a persistent identification network and other essential services. We have applied our concepts of transient connectivity and active persistence validation and translation to enable seamless mobility for IP devices like servers, laptop computers and PDAs, and non-IP devices as well, like Bluetooth enabled terminals. Our approach allows clients running regular IPv6 or IPv4 to communicate directly with Mobile Entities as the MEs roam. Communication is achieved regardless of the MEs current IP association or even the fact that they might or might not be able to establish IP connections at all. We effectively enable IPv4 systems and Bluetooth enabled devices to have an IPv6 presence that is persistent in nature. Mobility and communication across heterogenous network structures are the main application benefits of our work.

The rest of the paper is organized as follows: Section 2 introduces the problem of host mobility and identification, and presents some current proposed solutions to the problem.

[1]Digital Entities and Mobile Entities are used interchangeably throughout this paper.

Section 3 overviews some essential concepts of transient communications exploited throughout this paper. Section 4 goes through the implementation of some of these components, essentially Agents and gateways. Besides, we present the demonstrations that were conducted on top of the implemented framework to assert the usefulness of the underlying concepts in solving some current issues with the Internet.

## II. Mobility and Identification

Clearly, *mobility* and *host identity* are interrelated notions. Forwarding traffic to a mobile host can not be achieved within a network that identifies a host by a physical address. The IP address performed reasonably well as an identifier until truly mobile systems arrived. When host mobility is considered, the varying IP address can not be directly used for host identity. This translates into IP loosing any meaning of identity reference and consequently, it degenerates into a pure routing identifier. Various approaches have been adopted in solving the problem of host mobility on the Internet. Most of these approaches focus on inserting a level of indirection, whether at the routing level (network core) or at the end hosts (network edge) or through an overlay infrastructure. Some of these proposals address the Mobility issue by decoupling the host identity from the actual host location.

Since MEs tend to change attributes and state frequently, identifying these MEs becomes a complex task especially when the same Mobile Object can travel along different heterogeneous network structures and systems. This brings in the issue of identifying and locating digital entities. The basic assumption here is that Mobile Objects maintain their identity as their attributes change. To further illuminate the identification issue, we will present a brief review of some proposed solutions to host mobility.

On one hand, Mobile IP [5] and Nimrod [6], [7] are two purely routing solutions. Mobile IP assumes that a mobile host will maintain the same IP address on its home network. The home agent will handle the routing and make sure that all packets addressed to the mobile host's home IP address, will reach the moving host. This indirection elevates the home IP address from a physical identifier to a host identifier that is location independent. Nimrod [6], a next generation routing infrastructure, is another routing approach to mobility. It introduces the concept of global end point identifiers (EIDs). Nimrod uses EIDs to address packets where an EID is resolved into an actual attachment point by the routing infrastructure itself. Mobility in Nimrod [7] is enabled using the Dynamic Association Module (DAM) that takes care of updating the EID mappings in the routing infrastructure.

Peernet [8] is also a routing approach that separates identity from location by providing an alternative to IP addressing. In Peernet, location information is based on binary address trees that simplify routing. Hosts are expected to update their location information into specific servers so that peer hosts can find them.

On the other hand, some proposals introduce end-to-end approaches to mobility. Migrate [9] is a session layer approach to mobility that involves modifying the network stacks at the end hosts. Migrate uses domain names (or other naming systems) to identify and track hosts. Domain names are translated into actual IP addresses at run time, hence, separating host identities from their physical attachment points. The Host Identity Protocol (HIP) [10] uses a cryptographic key as the host identifier (HI) and introduces a new layer at the end host stack above the network layer for the translation of the HI into an actual IP address.

Separating physical addresses from identifiers is also approached through overlay networks on top of IP. The Internet Indirection Infrastructure (I3) [11] is an overlay network that utilizes rendezvous servers for clients to register triggers. I3 decouples the sending and the receiving actions where clients send traffic to the overlay and the latter takes care of forwarding the traffic to other interested clients that registered triggers in the system. Mobility in I3 is the direct consequence of Identification abstraction. This is achieved using the trigger *id* [11] that acts as a rendezvous point between the sender's and receiver's traffic.

The Session Initiation Protocol (SIP) [12] architecture is another overlay approach that can be efficiently reused to provide mobility [13], [14] with a readily available infrastructure. This avoids the redundancy introduced by simultaneous deployment with Mobile IP [5].

The proposal Hi3 [15] provides an attempt at making use of an overlay (I3) in conjunction with a secure direct end-to-end approach (HIP). This would allow for end-to-end traffic with HIP [10] (mobility, multi-homing, DOS resistance) that leverages an independent secure, integrated rendezvous infrastructure (I3 [11]) as an overlay to route the HIP control traffic (stability).

The aforementioned mobility overview is meant to show the different paradigms to identification in mobile environments that attempt at decoupling the identity of a host from its actual IP address. Home IP addresses, domain names, rendezvous ids and cryptographic keys are examples of endpoint identifiers that are meant to abstract the actual attachment point of an endpoint. We contend that any identification framework meant to address the same issues and operate in highly mobile and transient environments should at least support *scalability*, *persistence*, *abstraction* and *security*. This approach has been previously applied by Khoury [16] to SIP [12] systems using the Handle system to successfully abstract SIP user identities. We have chose the Handle System due to its intrinsic persistence, security and scalability. Besides, the fast resolution times and the distributed administration model that the Handle System offer, promote it as a candidate identification system suitable to operate in highly mobile environments. Other identification systems are either not truly persistent or constrained in terms of their update characteristics. With HIP, for example, the use of the attribute (public key in this case) as the actual identifier eliminates identity persistence across changes of the attribute. On another hand, using domain names to identify endpoints does not scale. This is because the Domain Name System (DNS) [17] has significant lag time [18], it does not naturally or securely implement distributed individual DNS entry administration, and it's architecture includes a single point of entry and starvation.

## III. Mobile Transient communication

This paper assumes all connections to be transient in nature but persistently identified. We in fact present a basic proof of concept implementation of some of the abstract notions introduced by our work towards a distributed persistent transient network architecture. These notions include a proposed Distributed Persistent Identification Network (D-PIN) as an overlay exposed by an expansion layer between the network core and the network edge. We have used the Handle System which is the largest persistent identification network that we are aware of at this time. Although, the Handle System architecture is not a fully distributed system which we call for in out final design, it still provides a universal basic access to registered digital objects [19]. It provides a secure global name service for digital objects over the Internet. In this paper, we succeed to apply the Handle System identification framework to networked Digital Entities, mainly devices and services. Each Digital Entity is assigned a *handle* i.e. a name that can be associated with a set of attributes. Some of these attributes can be location, permissions, administrators and state.

Simply put, decoupling the host identity from its attributes is implemented as follows: Our framework identifies all network devices and services using high level identifiers that are unique, global and persistent. The devices themselves, or respective delegates, are expected to maintain a valid binding between the device identifier and its present attributes (for example IP address, administrators, etc...). The freshness of the binding is implemented by agents that reside on the devices or on their delegates. These agents make sure that the identifier-to-address mapping is never stale rendering the device accessible even after it changes its attachment point. A new layer above the actual network layer that performs handle resolution into actual IPv6/v4 addresses is implemented at the endpoints (currently we are reusing DNS). Hence, a device is able to change its attachment point frequently, while remaining accessible through its persistent identifier. More details on this are presented in section IV. The abstraction and persistence of the identifiers, the scalability of the identification system and its inherent security are all features that the mobile transient internet architecture promotes.

Additionally, mobility is evolving into more complex forms that are imposing additional burdens on the current networks. The advent of dual mode phones (WiFi/Cellular), for example, compels the network to support not only the ability of the device to change attachment points, but also its ability to use different communication environments and identification mechanisms. This paper addresses the issue through the Persistent Coordinated Translation (PCT) gateways that enable protocol and service translation in addition to communication mechanism translation.

### A. Digital Entities and Persistent Identifiers

This section presents a brief overview of the Distributed Persistent Identification Network (D-PIN) that we postulate. Although we have used the Handle System [4] due to its current availability; a complete deployment of the postulated network would help exploit the full potential of our work. D-PIN is a successor to the current Handle System Implementation. It has all the characteristics of the Handle System. It allows for Mobile Objects to establish persistent identity over heterogeneous network structures. *Identity persistence* is a direct consequence of the separation of the identifier from its attributes. This is achieved using a secure global name service that guarantees the association of the identifier with its respective attributes over distributed communication [20]. *Security* is another crucial property of the system. The system will act as a certification authority assuring that attributes of the name/reference are securely transferred between the communicating ends. So, the naming system allows for secure name resolution and administration in a distributed fashion making it suitable to operate in highly mobile environments. Mobile Objects are identified irrespective of the communication environment they interact with. *Scalability* is also essential and is part of the system design and is accounted for in D-PIN.

### B. Agents

As mentioned earlier in the paper, Agents are needed to maintain the freshness of the system. Agents may be stationary or mobile. An example of static agents are update and infrastructure agents. Update agents are typically bound to a Mobile Entity and stay with it. This type of agent makes sure that the *handle* mappings of the Mobile Objects are never stale. Infrastructure agents reside on PCT Gateways and assist in disseminating and updating information within the system. On the other hand, Mobile Agents perform some crucial tasks like logical coordination and routing carrying data and instructions across networks. This paper will be concerned solely with non mobile agents. The coordination of these agents with the persistent naming system D-PIN allows for mobility and persistence in communication as we shall see in section IV.

### C. Persistent Coordinated Translation (PCT) Gateways

The concept of the PCT Gateway is essential to the harmony of the overall network infrastructure. We distinguish between two kinds of PCT Gateways: Edge PCT (E-PCT) and Interface PCT (I-PCT).

*1) E-PCT Gateway:* This type of Gateway is intended to prolong the reach of the Internet by continuously redefining network edges. In the case where an ME has limited resources, these gateways act as super agents that manage the *handle* identifiers of the MEs and provide the MEs with basic network translation resources. E-PCT gateways implement the Open Device Access Protocol (ODAP)[21]. ODAP is a novel protocol destined to discover and associate certain Mobile Objects (Devices, Services) and the heterogeneous networks through them to the overall persistent infrastructure.

*2) I-PCT Gateway:* Interface PCT (I-PCT) gateways handle the migration of non-persistent instantaneous traffic to transient persistent networks and vice versa. They are also responsible for interfacing current technologies with the persistent resources and facilities of the transient infrastructure. The agents residing on these gateways perform on the fly protocol translations such as DNS-to-Handle as well as application specific implementations.

## IV. Implemented Framework

This section details our implementation of some basic components that work together to deliver part of the transient network architecture functionality. The application we are interested in deals mainly with host mobility and heterogeneous communications. For the remaining part of this paper, we use the term MEs to refer to network devices and services. A complementary application by Khoury [16] shows how *users* become MEs in a *Voice-over-IP* framework while roaming across SIP [12] domains. The users are identified by *handles* which leverages identity persistence at the service level.

The Handle System acts as a global resolution/indirection platform that abstracts the IP specifics from devices. Our implementation associates non-IP, and in general deep web, (Bluetooth devices, robots,...) resources with global IPv6 or IPv4 addresses. This makes them accessible over the internet just as traditional IP enabled devices are. IP enabled devices will be directly accessed through their global IPv6/4 addresses, whereas, non IP devices will make use of an Interface PCT gateway that will associate global IPv6/4 addresses with them abstracting the lack of IP functionality. Communication with the Handle System, mainly resolution and administration, uses the Handle System protocol [3]. The gateway deployments will therefore allow for communication between devices residing on heterogeneous networks, while the persistent identification framework will foster mobility as we shall see.

A sketch of the overall implemented framework is depicted in Figure 1. It shows a particular implementation that exploits the notions of IP mobility and communication across heterogenous networks. The implementation reflects a house deployment of an E-PCT gateway that exposes home appliances, robots (non-IP devices) to the global Internet by binding them to global IPv6 addresses. A client on the Internet is now able to communicate with these devices and access the services they provide. We will now go through the details of the components and protocols involved in this implementation.

### A. Components and Protocols

*1) Persistent Identification Network and Agents:* We use current implementation of the Handle System [1], [2], [3], [4] to identify MEs. MEs therefore posses unique, global and persistent *handles*. Device *handles* map locally depending on their communication medium into physical addresses. So, for example, the *handle hdl/laptop* of a laptop computer that is connected to the Internet, can map into a global IPv6/4 address. Service *handles*, on the other hand, can map into service specific identifiers which could reside inside the operating system they are set to operate in. For example, the *handle hdl/UUID* could map into a specific service running on a Bluetooth device and identified with a certain UUID [22]. As another example, the *handle hdl/sipuser* of a SIP [12] user, can map into the actual SIP URL (domain binding) on which that user is reachable [16]. The attributes that the *handle*s map to are volatile in nature; however, the *handle*s themselves are persistent.

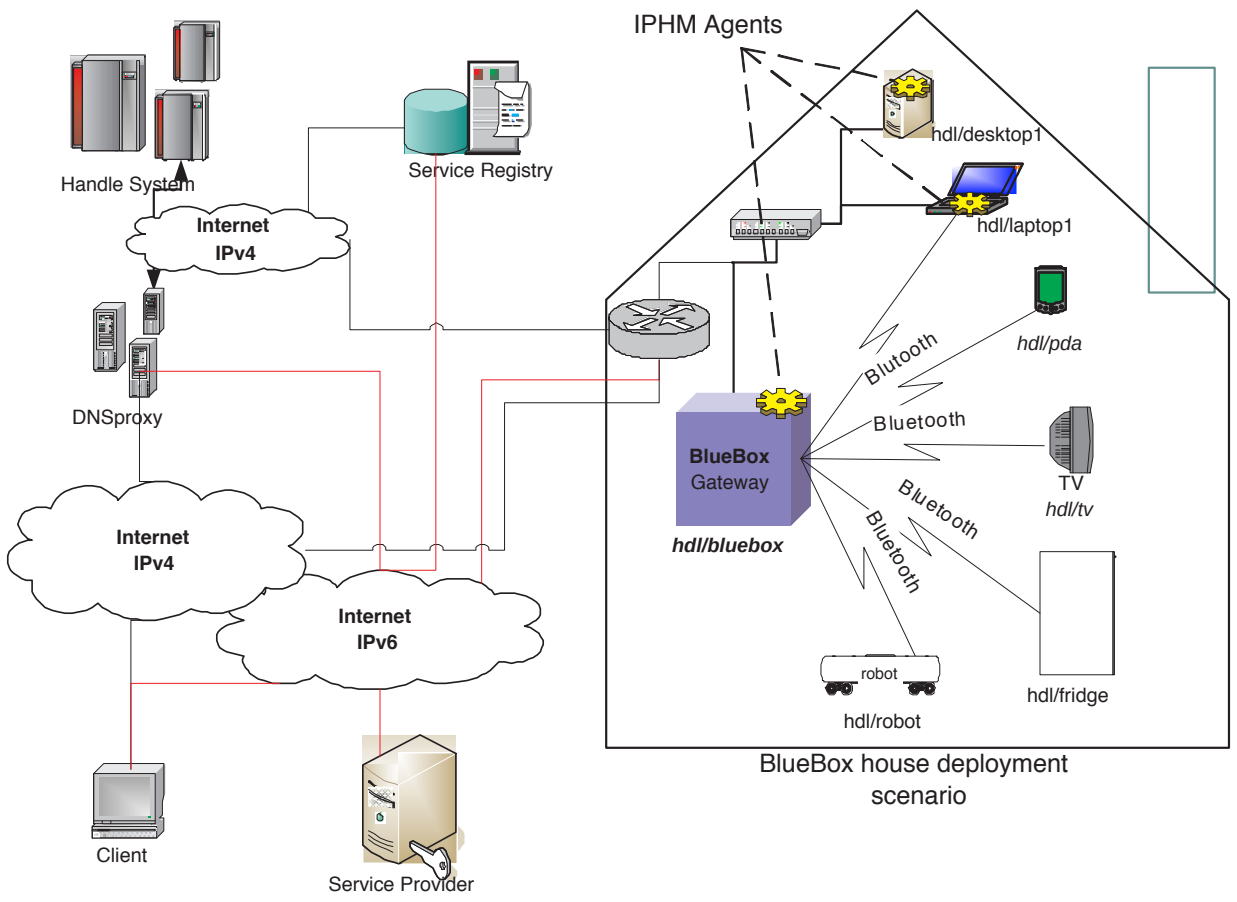

Fig. 1. A reference sketch of the implemented framework.

The *handle* thus becomes independent of the resource's location, ownership and other state information. So, changing a resource's location, for example, will not break the *handle* itself as long as the new location is updated in the *handle*. This raises the issue of updating the *handle* attributes according to administrative privileges. One type of stationary agents that we implemented, the IP Handle Monitor (IPHM), makes sure the *handle* bindings are never stale. In Figure 1, the IP devices *hdl/laptop1*, *hdl/desktop1* and the gateway *hdl/bluebox* run an IPHM agent allowing them to change their network attachment point while still being accessible through their *handle*s.

The implemented IPHM is a relatively simple software component that runs on network devices with abundant resources (laptops) or limited resources (PDA's), and monitors the device's attachment point. It is fed with the information about the particular device *handle* as well as the certificates needed to administer the *handle*. The currently implemented IPHM will detect any changes to the device's active IPv6/4 address and use this information to update the respective attributes for the particular *handle* in the Handle System. Perhaps the best way to describe the IPHM functionality is through the use of a flowchart. We distinguish two fairly simple algorithms that the IPHM currently executes. The first, Fig. 2, is responsible for device authentication with the Handle System. The second algorithm, Figure 3, takes care of IP administration. The notation <> in the flowcharts refers to dynamic parameters. Currently the algorithm prefers to bind global IPv6 addresses with device *handle*s. In the case where no global IPv6 address is detected, the *handle* is bound to the IPv4 address. As far as this paper is concerned, IPv4 addresses are assumed to be public.
The IPHM is implemented in Java. It runs on Windows and Linux Operating Systems. Besides, we have implemented a J2me version of the IPHM that runs on limited resource devices, mainly PDA's (Linux, Windows Mobile 2003). The J2me version was tested on the Zaurus SL600 (Linux) PDA and the Siemens SX66 (Windows Mobile 2003) PDA. Note that dual stack (ipv4/ipv6) network connectivity was used during deployment allowing resources to be accessed either through a public ipv4 address or preferably a global ipv6 address.

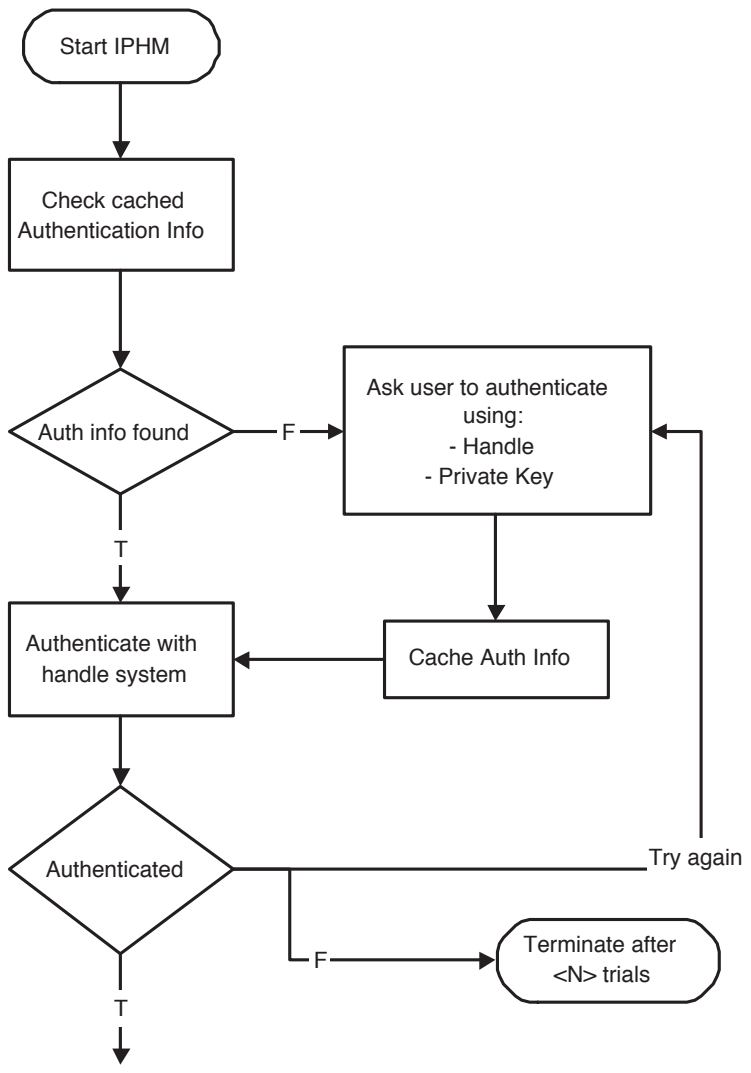

Fig. 2. IPHM Device Authentication algorithm

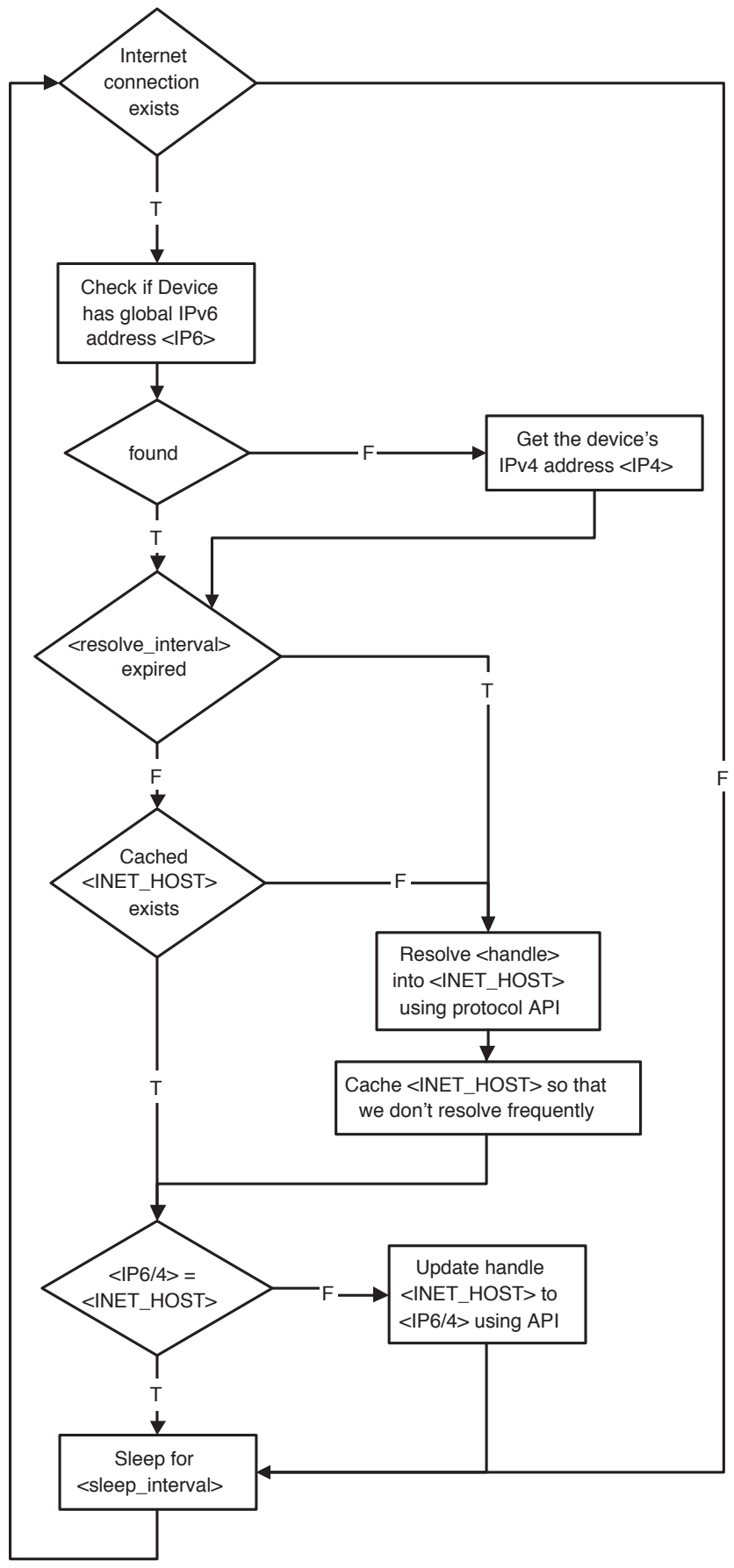

Fig. 3. IPHM IP Admin algorithm

*2) Open Device Access Protocol - ODAP:* Briefly, ODAP [21] is a protocol targeted at the discovery of nearby network devices and the association of these discovered devices with an IP address and an indirection global address. The protocol operates in push and pull modes. In the pull mode, a gateway running the ODAP, polls devices in its vicinity for the services they wish to advertise. In the push mode, the devices will initiate service registration requests with a gateway in their vicinity. The protocol's interface currently allows for four distinct operations including *Device Discovery*, *Service Listing*, *Service Implementation* and *Device handover preparation*. The behavior of these functions depends on the operational mode of the protocol.
We have implemented a basic version of the ODAP protocol interface operations on the *BlueBox* E-PCT gateway discussed in section IV-A.3.a.

*3) PCT Gateways:*

*a) E-PCT:* To illustrate the E-PCT type gateway, we have implemented a stand-alone product which we call the *BlueBox*. The *BlueBox* implements the Open Device Access Protocol. It allows devices residing on private networks that either posses a public/private IP, or do not posses an IP at all (e.g. Bluetooth devices, sensor network components, home appliances,...) to be exposed by associating the global IPv6/v4 addresses of these devices with their particular *handle*s. The actual IPv6/v4 addresses of these devices are associated with the *BlueBox* that will later on, either route or translate the traffic to the devices. The implemented gateway uses Linux as its operating system and communicates on two different interfaces. The Bluetooth interface is used for internal communication with Bluetooth devices. Externally, the gateway uses Ethernet/WiFi to connect to the Internet. Bluetooth is used to illustrate a non-IP environment. The ODAP is implemented in pull mode. The *BlueBox* runs an IPHM that monitors its IP address keeping its *handle* binding updated. A basic version of the ODAP is implemented on the *BlueBox* PCT Gateway in Java. The protocol interface operations are set to communicate over HTTP for demonstration purposes. The gateway receives client requests over HTTP and uses Bluetooth internally to communicate with the Bluetooth devices.
We will discuss the ODAP interface operations implemented on the *BlueBox* in light of the framework of Figure 1, where a client is trying to access the services provided by some devices residing behind the *BlueBox* gateway:

- *Device Discovery* As the gateway receives a client request for Device Discovery, it starts scanning for Bluetooth

devices in its vicinity. It then authenticates these devices and associates each with a global *handle* partly constructed of the device's Bluetooth MAC address. Note that a special algorithm is used to compute the *handle* of a Bluetooth device and verify it globally. This step is part of Mobile Device Authentication Protocol MDAP (protocol still under research), a protocol used to authenticate passive and active ODAP clients. It is worthwhile noting that presenting the authentication algorithm with a device (Bluetooth device in this case) will always yield the same unique *handle* for that device. A valid global IPv6/4 address from the gateway's pool of addresses is then associated with the computed *handle*. The *handle* is then either updated (*handle* already exists in the Handle System meaning that the device was previously registered with some gateway) to reflect the new IP address, or created (first time device registers with a gateway). At this point, the gateway can inform the requesting client of the online devices which are then listed as a set of handles.

- *Service Listing* The client then issues a Service Listing request for a specific Bluetooth device. The gateway in turn polls that device for the services/data the latter is advertising. The gateway generates the respective *handle*s of these services i.e. the gateway maps the services/data identifiers to their global *handle*s. Note that in the Bluetooth case, services are identified with unique UUIDs [22] that are part of the global service *handle*. The services for the device are then listed.
- *Service Implementation* If the service is not found locally by the gateway, the gateway resolves the service *handle* to locate the code required to implement the service. This last step involves the following sequence of events: The service handle is used to query a service registry. This query, which is contextualized to the particular characteristics of the gateway, is sent to the registry as an actionable request. The service registry points the gateway to a specific service provider site containing the actual signed code that fits the gateway's needs. This process involves secure implementation of the service which is itself identified by a persistent identifier.
  The mechanism depicted above for supporting service implementation is referred to as delegated implementation i.e. the gateway implements the service and it exposes a high level interface (web interface for example) to the client. We have implemented this approach on our *BlueBox*. On the other hand, service implementation can be done through encapsulation where the client encapsulates Bluetooth commands in IP and the gateway decapsulates these commands without implementing the service locally.
- *Device Handover Preparation* The gateway sends keep-alive verification messages to the devices to verify their presence. Upon failure to verify device presence for a defined interval of time, the device IP cleared and made available in the gateway's pool.

*b) I-PCT:* We have implemented an I-PCT gateway responsible for protocol translation, specifically, Handle-to-DNS. We refer to this gateway as the Handle-DNS proxy (HDP). HDP is a modified DNS server that communicates using the bind protocol and implements extra functionality allowing it to associate canonical names and aliases inside its particular naming zone with *handle*s. HDP will therefore resolve canonical names inside its naming zone using the Handle System and will, in addition, allow any common DNS server to resolve DNS entries in the format: <handle>.[DNS proxy domain] to the actual value of the INET_HOST attribute of that particular *handle*. Figure 5 shows how the HDP is integrated with the current DNS infrastructure.

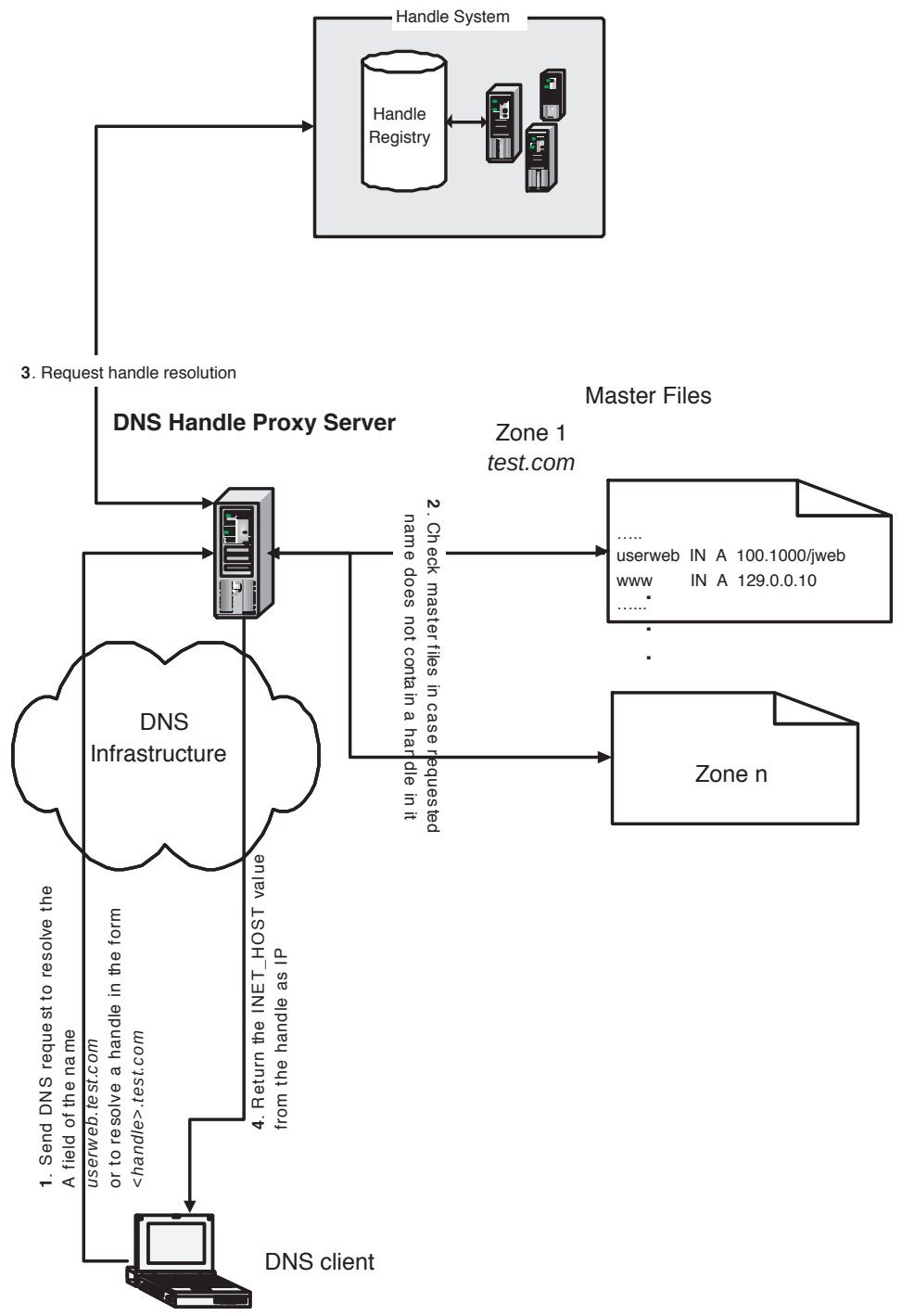

Fig. 4. Handle-DNS Proxy

Figure 4 shows how a client requests *handle* resolution just the same way it resolves traditional DNS names. The HDP checks for the canonical name of the DNS query in its zone files and if not found, assumes the name is a *handle* and tries to resolve it using the Handle System. The host names in the zone files can be associated with either static IP addresses (*www* in Fig. 4), or *handle*s (*userweb* in Fig. 4). In the latter case, an additional step is required to resolve the *handle*.

The significance of this approach becomes evident when the host IP changes frequently i.e. mobile host. If the host device is running an IPHM, the IP address of that device will be automatically reflected in the Handle System and no administrative changes need be implemented in the zone files. We implemented a global DNS proxy which accepts requests from any client and acts as mediator between DNS and the Handle System. A request to this server is a normal DNS query with query *name* having the following format: <name> = <hostname | handle> ".proxy.domain". Note here that the Global HDP will assume that the name part before "proxy.domain" is a *handle* if it is not a host name in the zone files and if obeys the handle naming syntax [1].

Possible values of DNS query names are "2118/resource1.dns.handle.net" or "userweb.dns.handle.net". The latter turns out to be the same as "100.1000/jweb.dns.handle.net" as depicted in Figure 4. DNS Clients (e.g. nslookup) should allow query names to contain the special characters used in the *handle* namespace like ascii '/' (0x2F). Refer to [1] for a detailed description of the *handle* namespace. Since Internet hostnames can not contain the '/' character [23], we replaced it with the '$\sim$' (0x7E) character for testing purposes. It is worthwhile noting that changing the IP address of the *handle* is done securely using PKI authentication by the *handle* administrator (IPHM in this case). Consequently, the simultaneous functionality of the IPHM, discussed in the previous section, and the HDP allows ubiquitous access to networked resources using the current network infrastructure. HDP is implemented in Java. It is currently running on a Fedora Core 4 OS.

### B. Demonstrations

The following demonstrations clarify how mobility and heterogeneous communication are achieved.

*1) IPHM and DHP:* The functionality of the IPHM is demonstrated on two kinds of devices with different computing resources, laptops and PDA's. In both cases the IPHM is fed into the device with the respective authentication information. When the IPHM starts first, it tries to authenticate with the Handle System asking the user for the *handle* and the private key.

When the *handle* is successfully authenticated, the authentication information is encrypted and cached internally so that subsequent initiation of the IPHM (e.g. rebooting the device) uses the cached authentication info and does not require further user authentication. A similar scenario takes place when the IPHM starts on the PDA. For our demonstrations, a Zaurus SL6000 linux PDA WiFi enabled is used. As the device connects to a different network, the IPHM automatically detects and updates the new IP address binding.

If the device supports dual stack (IPv6/IPv4) and obtains an IPV6 address, then the IPv6 address is reflected in the *handle*. However if the device is only IPv4 enabled, the IPv4 address is reflected by the *handle*. In the event that the IPv4 address is globally routable no further action is needed. If the IPv4 address is private then NAT traversal is required. The NAT traversal, that is also part of our research, is beyond the scope of this paper. The IPHM is dispatched as a light-weight process that runs in the background of the device and is transparent from the user. It is anticipated to become part of the operating systems.

Further, a web server was installed on both the laptop and the PDA. With the help of the Handle-DNS Proxy described earlier, we are able to access the web server using the *handle* of the device. Thus accessing `http://hdl~pda.proxy.domain/` in a browser always directs us to the web server running on the zaurus PDA (Note here that the zaurus PDA *handle* is hdl/pda). This eliminated the need to update DNS entries should the PDA connect to another network. Persistent identification is thus demonstrated. Most recent browsers (IE, FireFox) support IPv6 and prefer it by default. So, the browser will attempt to resolve the *handle* into an IPv6 address first through the HDP.

This demonstration aimed at presenting the mutual functionality of the IPHM and the HDP and their effective role in allowing IP mobility of connected devices over the current internet.

*2)* BlueBox*:* This section demonstrates the functionality of the *BlueBox* as well as the Open Device Access Protocol ODAP. Figure 5 illustrates the communication flow and components involved in the implemented framework.

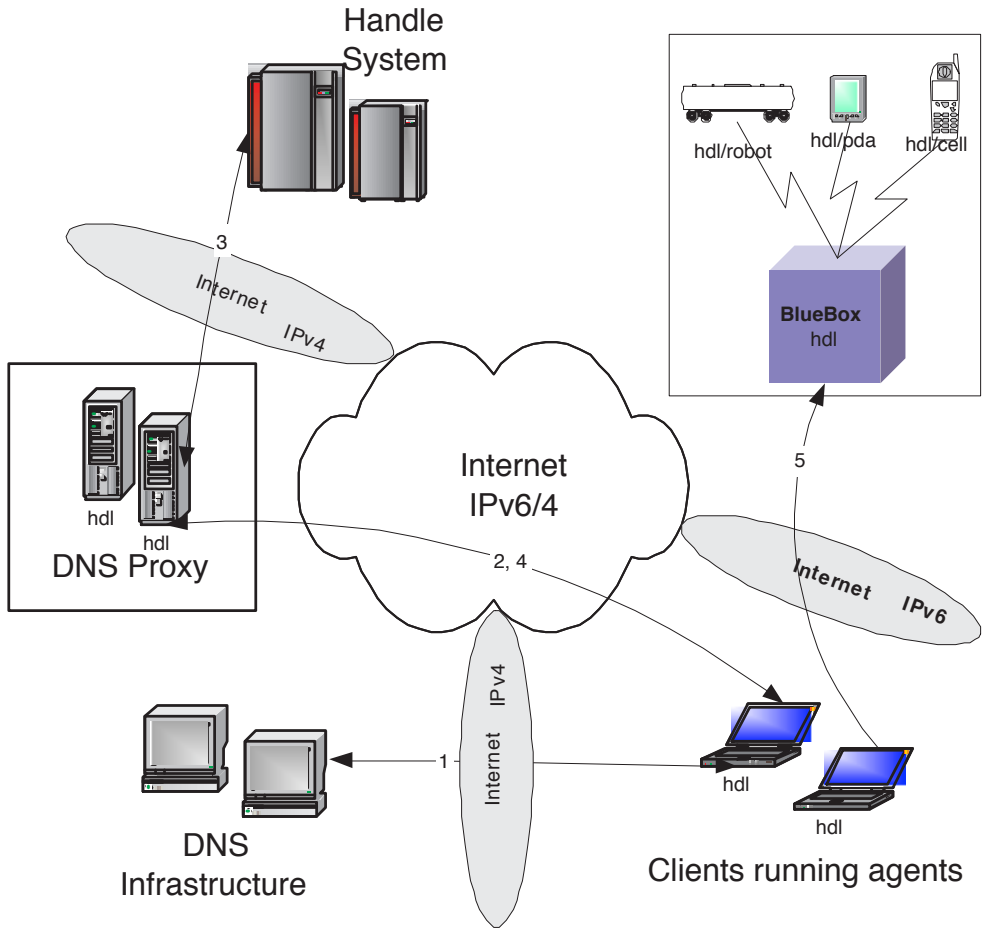

Fig. 5. Implemented Framework

The gateway or *BlueBox* is a linux machine. Recall that it has two communication interfaces, Bluetooth and WIFI/Ethernet interfaces. The gateway allows clients to access it through an HTTP interface. Note that clients are either IPv4 only, IPv6 only or dual stack clients and so is the gateway. However, throughout this demo, both the clients and the gateway are dual stack and use IPv6. Note also that the demonstration version of ODAP runs on top of HTTP and consequently the gateway runs an HTTP server (tomcat 5.5.9 is used). The future version of ODAP is anticipated to implement its own communication protocol.

After booting the gateway and connecting it to the internet, some Bluetooth devices are placed in its vicinity. These devices include a mobile phone, a PDA and a Robot, all of which have their Bluetooth interfaces turned on and allow other Bluetooth devices to see them and access their services (Figure 5). A client browser sends an HTTP request to `http://hdl~bluebox.proxy.domain/` which is resolved by the browser into the IPv6 address of the gateway (A global HDP is running on the domain proxy.domain) as depicted in Figure 5 (arrows 1,2,3,4).
The client issues a *Device Discovery* request. The gateway in this case lists the discovered devices in the client's browser. At this point, the client sends a *Service Listing* request for the robot. The gateway sends back a list of the services exposed by the robot, mainly the robot control service. Client requests service implementation of the robot control service. Note that this service is made available using the LabVIEW software [24] which allows a web interface for controlling the robot. The gateway sends back a URL to client pointing the robot control web interface `http://hdl~robotcontrolservice.proxy.domain` allowing the client to control the robot. In the meantime the gateway is always sending keep-alive messages to the devices to verify their presence and maintain their global IPv6 binding.
We also demonstrated communication between an IPv6 device and another IPv4 device where the gateway handled the translation. The protocol itself is abstracted from the communicating ends.

## V. Future Work and Enhancements

One of the limiting factors of the implemented framework is the overhead needed to resolve *handle*s. Since *handle*s can identify highly mobile devices, caching is not a solution. This encourages future implementations of the Handle System that would allow for significantly faster resolutions. Besides, the transient internet architecture postulates a fully Distributed overlay of the Persistent Identification Network (D-PIN) and proposes an advanced implementation of the Handle System in a structured P2P fashion that would make it more reliable. Mobile agents that handle routing based on persistent identification is one of the topics we are currently addressing.
This paper is simply a mobility application to the transient network architecture. Future papers are intended to address additional paradigms such as the Distributed PIN implementation, persistent identifier routing, etc. Current research is focusing on implementations of the Green Network, a new network that would introduce persistence at the network level to all transient communications. Additional work is underway in the service implementation side to enable mobile signed code implementation. This feature will allow the PCT gateways to resolve the code necessary to implement a particular service once it has been mapped into its particular persistence identifier. This feature which goes beyond the postulations of CORBA [25] would enable full mobility and automatic service functionality updates.

## VI. Conclusion

In this paper we have demonstrated an application implementation of a transient network architecture. We have established the usefulness and applicability of the Persistent identification network PIN to identify and locate network devices and services. Demonstrations used the current implementation of the Handle System. We have also proved the notions of mobility and persistence in communication across heterogeneous networks using the indirection framework.

## VII. Acknowledgements

We would like to thank all our sponsors.

## References

[1] S. Sun, L. Lannom, and B. Boesch, "Handle system namespace and service definition," RFC 3651, November 2003.
[2] S. Sun, L. Lannom, and B.Boesch, "Handle system overview," RFC 3650, November 2003.
[3] S. Sun, S. Reilly, L. Lannom, and J. Petrone, "Handle system protocol (ver2.1) specification," RFC 3652, November 2003.
[4] "The handle system," http://www.handle.net.
[5] C. E. Perkins, "Ip mobility support for ipv4," RF 3220, January 2002.
[6] I. Castineyra, N. Chiappa, and M. Steenstrup, "The nimrod routing architecture," Internet Engineering Task Force RFC 1992, August 1996.
[7] R. Ramanathan, "Nimrod mobility support," Internet Engineering Task Force RFC 2103, February 1997.
[8] J. Eriksson, M. Faloutsos, and S. Krishnamurthy, "Peernet: Pushing peer-to-peer down the stack." IPTPS 03, February 20,21 2003.
[9] A. C. Snoeren and H. Balakrishnan, "An end-to-end approach to host mobility," in *Sixth Annual ACM/IEEE International Conference on Mobile Computing and Networking*, August 2000.
[10] R. Moskowitz, P. Nikander, and P. Jokela, "Host identity protocol," Internet Draft, work in progress, draft-moskowitz-hip-09.txt, IETF, 2004.
[11] I. Stoica, D. Adkins, S. Zhuang, S. Shenker, and S. Surana, "Internet indirection infrastructure," in *ACM Transactions on Networking*, vol. 12, no. 2. IEEE, April 2004.
[12] J. Rosenberg, H. Schulzrinne, and etal., "Sip: Session initiation protocol," RFC 3261, June 2002.
[13] E. Wedlund and H. Schulzrinne, "Mobility support using sip," 2nd ACM/IEEE International Conference on Wireless and Mobile Multimedia, Seatle, WA, August 1999.
[14] H. Schulzrinne and E. Wedlund, "Application-layer mobility using sip," in *Service Portability and Virtual Customer Environments*. IEEE, December 2000, pp. 29–36.
[15] P. Nikander, J. Arkko, and B. Ohlman, "Host identity indirection infrastructure (hi3)."
[16] J. Khoury, H. Jerez, and C. Abdallah, "H-sip: An approach to inter-domain sip mobility," 2006, pre-print available at https://dspace.istec.org/handle/1812/53.
[17] P. Mockapetris, "Domain names implementation and specification," IETF Networking Group RFC 1035, STD13, November 1987.
[18] C. Huitema and S. Weerahandi, "Internet measurements: the rising tide and the dns snag." in *Proceedings of the 13th ITC Specialist Seminar on IP Traffic Measurement Modeling and Management*, ser. IPseminar. Monterrey, CA, USA: ITC, September 18-20 2000.
[19] R. Kahn and R. Wilensky, "A framework for distributed digital object services," Internet Whitepaper http://www.cnri.reston.va.us/k-w.html, January 1995.
[20] S. Sun, "Establishing persistent identity using the handle system," Tenth International World Wide Web Conference, May 2001.
[21] H. Jerez and J. Khoury, "Open device access protocol (odap)- working whitepaper," http://hdl.handle.net/4263537/5023, May 2006.
[22] "Bluetooth technology," http://www.bluetooth.org.
[23] K. H. etal., "Dod internet host table specification," RFC 952, October 1985.
[24] "Labview software from national instruments." http://www.ni.com/labview/.
[25] "The common object request broker: Architecture and specification," http://www.omg.org/library/c2indx.html.